\pdfoutput=1
\documentclass[a4paper,twocolumn]{autart}

\usepackage{amssymb}
\usepackage{graphicx}
\usepackage[fleqn]{amsmath}
\usepackage{url}
\usepackage{IEEEtrantools}
\usepackage{eucal}

\newtheorem{theorem}{Theorem}[section]
\newtheorem{remark}[theorem]{Remark}
\newtheorem{lemma}[theorem]{Lemma}

\newtheorem{problem}[theorem]{Problem}
\newtheorem{definition}[theorem]{Definition}

\DeclareMathOperator{\cone}{Cone}
\DeclareMathOperator{\ideal}{Ideal}
\DeclareMathOperator{\monoid}{Monoid}

\begin{document}

\begin{frontmatter}

\title{Reference Dependent Invariant Sets: Sum of Squares Based Computation and Applications in Constrained Control\thanksref{footnoteinfo}}

\thanks[footnoteinfo]{This research has been funded by the Fonds National de La Recherche Scientifique (FNRS)- Mandat d'impulsion scientifique (MIS) ``Optimization-free Control of Nonlinear Systems subject to Constraints" Ref. F.4526.17 and the Ministerio de Econom\'ia y Competitividad of Spain under project DPI2016-76493-C3-1-R co-financed by European FEDER Funds. This paper was not presented at any IFAC meeting.}

\author[ULB]{Andres Cotorruelo\corauthref{Corres}}\ead{acotorru@ulb.ac.be},
\author[SL]{Mehdi Hosseinzadeh}\ead{mehdi.hosseinzadeh@ieee.org},
\author[US]{Daniel R. Ramirez}\ead{danirr@us.es},
\author[US]{Daniel Limon}\ead{dlm@us.es},
\author[ULB]{Emanuele Garone}\ead{egarone@ulb.ac.be}
\corauth[Corres]{Corresponding author.}

\address[ULB]{Service d'Automatique et d'Analyse des Syst\`{e}mes (SAAS), Universit\'{e} Libre de Bruxelles (ULB), Brussels, Belgium}
\address[SL]{Department of Electrical and Systems Engineering, Washington University in St. Louis, Missouri, USA}
\address[US]{Departmento de Ingenier\'ia de Sistemas y Autom\'atica, Escuela T\'ecnica Superior de Ingenier\'ia, Universidad de Sevilla, Seville, Spain}


\begin{abstract}
The goal of this paper is to present a systematic method to compute reference dependent positively invariant sets for systems subject to constraints. To this end, we first characterize these sets as level sets of reference dependent Lyapunov functions. Based on this characterization and using Sum of Squares (SOS) theory, we provide a polynomial certificate for the existence of such sets. Subsequently, through some algebraic manipulations, we express this certificate in terms of a Semi-Definite Programming (SDP) problem which maximizes the size of the resulting reference dependent invariant sets. We then present the results of implementing the proposed method to an example system and propose some variations of the proposed method that may help in reducing the numerical issues of the method. Finally, the proposed method is employed in the Model Predictive Control (MPC) for Tracking scheme to compute the terminal set, and in the Explicit Reference Governor (ERG) scheme to compute the so-called Dynamic Safety Margin (DSM). The effectiveness of the proposed method in each of the schemes is demonstrated through a simulation study.
\end{abstract}

\end{frontmatter}

\section{Introduction}
The relevance of positively invariant sets lies in their numerous applications \cite{blanchini1999set}. More precisely, given an autonomous dynamical system, a subset of the state space is said to be \textit{positively} invariant if, assuming it contains the state of the system at some time, it will also contain it in the future.

Invariance is particularly important in the control of dynamical systems subject to constraints. In particular, in Model Predictive Control (MPC) \cite{Mayne2000}, the use of an invariant set as a terminal constraint is typically used to ensure stability and recursive feasibility. Additionally, invariant sets are at the basis of Reference Governor (RG) \cite{garone2017reference} approaches, of the recently introduced Explicit Reference Governors (ERG) \cite{Nicotra2018}, and of multimode regulators for constrained control \cite{kolmanovsky1997multimode}.

The computation of positively invariant sets has been the object of many research works. In \cite{haimovich2010componentwise}, the authors compute polyhedral invariant sets for switched linear systems. In \cite{pluymers2005efficient}, a method to compute a polyhedral invariant set for linear systems with polytopic uncertainty is presented. For what concerns input saturated systems, in \cite{alamo2005improved} the authors present a method to compute invariant ellipsoids, and in \cite{alamo2006new} a novel concept of invariance for saturated systems is introduced. For an extensive survey on research in invariant sets and their usage in control, the reader is referred to \cite{Blanchini2008}.

The majority of the work present in the literature focuses on finding domains of attraction for a single point of equilibrium. However, in the realm of reference tracking, the required invariant sets need to be centered around any admissible point of equilibrium, and thus become parameterized in the reference. These sets are much less straightforward to be computed explicitly.

To the best of our knowledge, the first instance of this concept dates to the Maximal Output Admissible Sets \cite{gilbert1991linear}. These sets are defined as the set of pairs of initial states and references such that the trajectory of the system fulfills the constraints when said reference is kept constant. For discrete time systems, some polyhedral representations of these sets can be computed for linear \cite{kolmanovsky1995maximal} and nonlinear systems \cite{hirata2008exact}, or characterized using Lyapunov arguments \cite{gilbert1999set}.

Due to the link between the Maximal Output Admissible Sets and the Dynamic Safety Margin (DSM) in the ERG framework, this line of research has seen a recent increase in activity. In \cite{garone2018explicit}, a closed form solution of the optimal DSM for linear systems subject to linear constraints is provided. In \cite{Hosseinzadeh2018}, the authors present a method to estimate the DSM implicitly in the case of convex Lyapunov functions and concave constraints. A method able to work with unions and intersections of concave constraints has been presented in \cite{Hosseinzadeh2019Letter}. 

A promising tool to tackle the computation of reference dependent invariant sets is Sum of Squares (SOS) programming \cite{parrilo2000structured}. In recent years the SOS framework has been used extensively to tackle invariance-related problems. For example, it was shown that it is possible to compute estimations of the basin of attraction of polynomial systems \cite{tan2008stability} and non-polynomial systems \cite{chesi2005domain}.

In this paper we propose to employ the SOS framework to compute reference dependent invariant sets for constrained systems. The main idea is to fix a polynomial Lyapunov function parameterized in the reference, and through SOS arguments compute the largest level set fully contained in the constraints. The effectiveness of the proposed methods is showcased with two applications: a discrete time system controlled with an MPC for Tracking, and a continuous time system controlled with an ERG.

The rest of this paper is organized as follows. Section \ref{sec:PS} states the problem. In Section \ref{sec:method} we provide a polynomial certificate for the existence of reference dependent invariant sets, along with an optimization problem to compute them. In Section \ref{sec:practical} some practical considerations that might aid in reducing the possible numerical issues are discussed. Two applications are presented in Section \ref{sec:app}: (i) characterizing the terminal set in the MPC for Tracking framework in Subsection \ref{sec:appMPC}, and (ii) providing a measure of safety in the  ERG framework in Subsection \ref{sec:appERG}. Some simulation results are also provided to show the effectiveness of the proposed method. Finally, in Section \ref{sec:conclusion} some conclusions and possible future research lines are presented.
\paragraph*{Notation:}
The set of polynomials with variables $x_1,\ldots,x_n$, whose coefficients belong to $\mathbb{R}$, is denoted by $\mathbb{R}[x_1,\ldots,x_n]$. We denote the set of all positive reals plus zero, and the set of all positive integers plus zero as $\mathbb{R}_{\geq0}$ and $\mathbb{Z}_{\geq0}$, respectively. For polynomials $p_j$, $j=1,\ldots,N$, we will use $\{p_j\}_{i=1}^N$ to denote the set $\{p_1,\ldots,p_N\}$. The set of all Sum of Squares polynomials with variables $x_1,x_2,\ldots,x_n$ is denoted by $\Sigma[x_1,x_2,\ldots,x_n]$. We denote the degree of a polynomial $p$ by $\partial p$. The $n$-dimensional identity matrix is denoted by $I_n$. The Jacobian matrix of a vector valued function $f$ is denoted by $\nabla f$. The gradient of a scalar function $V$ is denoted as $\nabla V$. For two sets $\mathcal{A},\mathcal{B}\subseteq \mathbb{R}^n$, $\mathcal{A}\ominus\mathcal{B}$ denotes the Pontryagin difference. We denote the $n$-dimensional ball of radius $\varepsilon$ by $\mathcal{B}_n(\varepsilon)\triangleq \{x\in\mathbb{R}^n : x^Tx\leq \varepsilon^2 \}$.
\section{Problem Statement}\label{sec:PS}
Let the system
\begin{equation}\label{eq:system_open_loop}
\delta x=\phi(x,u),
\end{equation}
where $\delta x$ is the successor state in discrete time or the time derivative in continuous time, $x\in\mathbb{R}^n$ is the state, and $u\in\mathbb{R}^m$ is the input. The system is subject to constraints in the following form:
\begin{equation}
x\in\mathcal{X}\,,~u\in\mathcal{U}\,,
\end{equation}
where $\mathcal{X}\subseteq\mathbb{R}^n$ and $\mathcal{U}\subseteq\mathbb{R}^m$ are simply connected sets with nonempty interiors.

\begin{figure}[t]
\centering
\includegraphics[width=7cm]{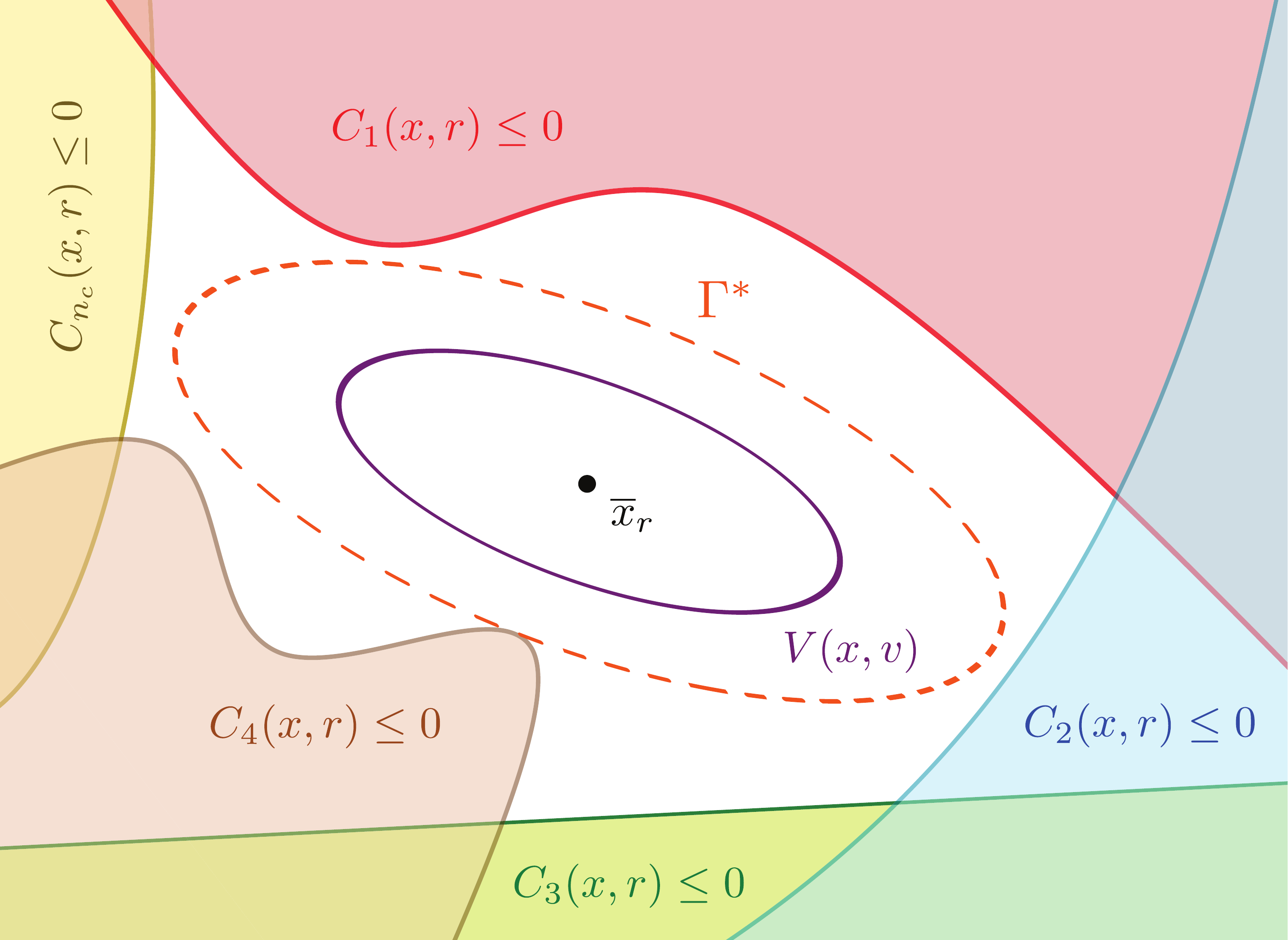}
\caption{Geometric illustration of the class of constraints under study.}\label{fig:GeoInt1}
\end{figure}

As often done in reference tracking problems, it is assumed that a control law $u=\kappa\left(x,r\right)$ is used to stabilize\footnote{The stabilization of unconstrained nonlinear systems is the subject of an extensive literature (\textit{e.g.}, \cite{Schmid2010,Ntogramatzidis2016}), and can be approached using a variety of available control tools.} the system, where $r\in\mathbb{R}^p$ is the reference signal. System \eqref{eq:system_open_loop} then becomes
\begin{equation}\label{eq:system}
\delta x=f\left(x,r\right),
\end{equation}
where $f\left(x,r\right)\triangleq \phi\left(x,\kappa\left(x,r\right)\right)\in\mathbb{R}[x,r]$ and the pair $\left(x,r\right)$ is constrained in the set $\mathcal{D}\triangleq\left\{\left(x,r\right):x\in\mathcal{X}\right.$, $\left.\kappa\left(x,r\right)\in\mathcal{U}\right\}$. Typically, this set can be expressed through a set of inequalities as
\begin{equation}\label{constraint}
\mathcal{D}=\{(x,r):c_i(x,r)\geq0,~i=1,\ldots,n_c\}.
\end{equation}
In this paper we assume that these inequalities are polynomial, \textit{i.e.} $c_i(x,r)\in\mathbb{R}[x,r],~i=1,\ldots,n_c$ (see \figurename~\ref{fig:GeoInt1}). For reasons that will be clearer later on, we also define the set of admissible references
\begin{equation}\label{eq:r_admissible}
\mathcal{R}\triangleq\{r\in\mathbb{R}^p:\overline{c}_i(r)\geq0,~i=1,\ldots,n_c\},
\end{equation}
where $\overline{c}_i(r)\triangleq c_i(\overline{x}_{r},r)$, with $\overline{x}_r$ as the equilibrium point of \eqref{eq:system} associated with a constant reference signal $r$ , \textit{i.e.}, $f(\overline{x}_r,r)=0$ in continuous time, and $f(\overline{x}_r,r)=\overline{x}_r$ in discrete time. Furthermore, $\mathcal{R}$ is assumed to be connected.

The purpose of this paper is to solve the following problem.
\begin{problem}{(Safe Reference Dependent Positively Invariant Sets)} Consider system \eqref{eq:system} where the pair $(x,r)$ is constrained in the set $\mathcal{D}$ as in \eqref{constraint}. Compute a family of sets parameterized in the reference $\mathcal{S}(r)$ such that
\begin{itemize}
    \item $\mathcal{S}(r)$ is invariant for system \eqref{eq:system} for any constant $r\in\mathcal{R}$, \textit{i.e.}, every trajectory of \eqref{eq:system} with initial condition $x(t_0)\in\mathcal{S}(r)$ and a constant reference $r\in\mathcal{R}$ is such that $x(t)\in\mathcal{S}(r)$ for all $t\geq t_0$,
    \item $\mathcal{S}(r)$ is fully contained in the constraint set $\mathcal{D}$ for every $r\in\mathcal{R}$.
\end{itemize}
\end{problem}
\begin{remark}
Contrary to regulation problems where a single invariant set centered around the origin is computed,
 in reference tracking we need to compute invariant sets centered around all possible points of equilibrium $\overline{x}_r$ associated with $r\in\mathcal{R}$.
\end{remark}
It is well known that one way to determine invariant sets is using Lyapunov level sets. In this paper we will assume, without loss of generality, that for each reference $r$, the corresponding equilibrium point of system \eqref{eq:system} denoted by $\overline{x}_r$ is globally asymptotically stable (see Remark \ref{rem:local} for local stability). We will also assume that a polynomial Reference Dependent Lyapunov Function (RDLF) $V(x,r):\mathbb{R}^n\times\mathbb{R}^p\rightarrow\mathbb{R}_{\geq 0}$ is known such that:
\begin{equation}\label{eq:Lyap_def}
\Bigg\{
\begin{aligned}
V(x,r)&>0\ \forall x\in\mathbb{R}^n\setminus \{\overline{x}_r\},\,V(x,r)=0 \Leftrightarrow x=\overline{x}_r\\
\delta V(x,r)&<0\ \forall x\in\mathbb{R}^n\setminus \{\overline{x}_r\},\, \delta V(x,r) = 0\Leftrightarrow x=\overline{x}_r
\end{aligned}.
\end{equation}
where $\delta V(x,r)$ denotes $\nabla V(x,r)f(x,r)$ in the continuous time case, and $V(f(x,r),r)-V(x,r)$ in the discrete time case.

It is easy to see that if the level set $\mathcal{S}_\Gamma(r)=\{x\in\mathbb{R}^n:V(x,r)\leq\Gamma(r),\,\Gamma(r)>0\},\,\ \forall r\in\mathcal{R}$ is fully contained in $\mathcal{D}$, then $\mathcal{S}_\Gamma(r)$ is a safe positively invariant set.

Typically it is useful to determine the largest possible safe invariant set. In the case of invariant sets based on Lyapunov functions, this corresponds to finding the largest bound $\Gamma^\ast(r)$ corresponding to the solution of the following optimization problem:
\begin{equation}\label{eq:optGammaStar}
\Gamma^\ast(r)=\left\{
\begin{array}{ll}
     &\max\Gamma \\
\text{s.t.}& \\
& \left\{x\in\mathbb{R}^n:V(x,r)\leq\Gamma\right\}\subseteq\mathcal{D}
\end{array}
\right.,
\end{equation}
for every $r\in\mathcal{R}$. Note that this optimization problem is parameterized in $r$ and that, except in a very simple case \cite{garone2018explicit}, its closed form parametric solution might be hard to compute and/or to handle.

In this paper we propose a systematic method to compute a good polynomial approximation of $\Gamma^\ast(r)$, denoted by $\hat\Gamma(r)\in\mathbb{R}[r]$ such that $\hat\Gamma(r)\leq\Gamma^\ast(r),\ \forall r\in\mathcal{R}$. Note that for any lower bound $\hat{\Gamma}(r) \leq \Gamma^\ast(r)$, the set $\mathcal{S}_{\hat{\Gamma}}(r)$ is a safe reference dependent invariant set. 

For the sake of simplicity, in the sequel we will compute one $\hat{\Gamma}(r)$ at a time. This is without loss of generality, since the multi constraint case can be expressed as the composition of single constraint cases as follows:
\begin{equation}\label{eq:multiconstraint}
    \Gamma^\ast(r) = \min_i \{\Gamma^\ast_i(r)\},~i=1,\ldots,n_c,
\end{equation}
where $\Gamma^\ast_i(r)$ is the value of $\Gamma^\ast(r)$ if only the $i$-th constraint $c_i(x,r)$ is considered.

\begin{remark}\label{rem:local}
In the case of local stability, it is possible to impose the additional constraint $\delta V(x,r)<0$ in \eqref{eq:optGammaStar}, which will yield the largest safe level set of the local RDLF.
\end{remark}
\section{Approximating the Largest positively Invariant Set via SOS Techniques}\label{sec:method}
In this paper we propose a method to compute a parameterized solution for \eqref{eq:optGammaStar} using SOS programming. This framework is able to tackle convex relaxations of non-convex optimization problems through polynomial optimization \cite{jarvis2005control}.

As usually done when working with the Krivine--Stengle Positivstellensatz (P-satz) \cite{papachristodoulou2005tutorial}, the first step is to express the conditions for an admissible solution of problem \eqref{eq:optGammaStar} in terms of set emptiness. In the case at hand, the condition that an admissible $\hat{\Gamma}_i(r)$ must fulfill is
\begin{equation}\label{eq:emptyset0}
    \left\{(x,r):V(x,r)\leq\hat{\Gamma}_i(r)\right\}\subseteq \mathcal{D}, ~\forall r \in \mathcal{R},
\end{equation}
where $V(x,r)$ is an RDLF as in \eqref{eq:Lyap_def}. It is straightforward to reformulate \eqref{eq:emptyset0} as the following set emptiness condition
\begin{equation}\label{eq:emptyset1}
\begin{aligned}
\left\{(x,r):r\in\mathcal{R},\,c_i(x,r)=0,\,V(x,r)>\hat{\Gamma}_i(r)\right\}=\emptyset.
\end{aligned}
\end{equation}
The next step is to apply the Krivine--Stengle P-satz \cite{tan2008stability}. Since in this formulation of the P-satz (See Appendix) the semialgebraic set required to be empty is described in terms of \textit{greater-than-or-equal-to}, \textit{not-equal-to}, and \textit{equals} operators, the set \eqref{eq:emptyset1} must be rewritten as\footnote{For brevity, we will omit the arguments of functions whenever there is no risk of confusion.}:
\begin{equation}\label{eq:emptyset}
\begin{aligned}
\left\{(x,r):\hat{\Gamma}_i-V\geq 0,\,V-\hat{\Gamma}_i\neq 0,\,c_i=0,\,\{\overline{c}_j\}_{j=1}^{n_c}\geq0\right\}.
\end{aligned}
\end{equation}
Note that, although we consider only one constraint for the computation of $\hat{\Gamma}_i$ (\textit{i.e.} $c_i(x,r)=0$), we still limit $r\in\mathcal{R}$ through $\overline{c}_1(r)\geq0,\ldots,\overline{c}_{n_c}(r)\geq0$. 

At this point, using the Krivine--Stengle P-satz and some standard algebraic manipulations, it follows that the set \eqref{eq:emptyset} is empty if there exist SOS polynomials $s_1,\ldots,s_{n_c+1}\in\Sigma[x,r]$, a polynomial $\bar{q}\in\mathbb{R}[x,r]$, and a non-negative integer $k$, such that
\begin{multline}\label{eq:SOSconditions0}
-s_0\cdot(V-\hat{\Gamma}_i)+(V-\hat{\Gamma}_i)^{2k}+\bar{q}\cdot c_i\\
-(V-\hat{\Gamma}_i)\sum_{j=1}^{n_c}s_j\cdot\overline{c}_j+s_{n_c+1}=0.
\end{multline}
If we impose $k=1$, $s_{n_c+1}=0$, and $\bar{q}=q(V-\hat{\Gamma}_i)$ where $q\in\mathbb{R}[x,r]$ it follows that, \eqref{eq:emptyset} is fulfilled if
\begin{equation}\label{eq:SOSconditions1}
(V-\hat{\Gamma}_i)(-s_0+(V-\hat{\Gamma}_i) +q\cdot c_i-\sum_{j=1}^{n_c}s_j\cdot\overline{c}_j)=0,
\end{equation}
which allows us to formulate the following sufficient condition
\begin{equation}\label{eq:SOSconditions2}
s_0=V-\hat{\Gamma}_i+q\cdot c_i-\sum_{j=1}^{n_c}s_j\cdot\overline{c}_j.
\end{equation}
Since $s_0$ is a SOS polynomial, \eqref{eq:SOSconditions2} implies that $\hat{\Gamma}_i$ is a lower bound of $\Gamma^\ast_i$ if there exist SOS polynomials $s_1,\ldots,s_{n_c}$ and a polynomial $q$ such that
\begin{equation}\label{eq:SOSconditions}
    V-\hat{\Gamma}_i+q\cdot c_i-\sum_{j=1}^{n_c}s_j\cdot\overline{c}_j\in\Sigma[x,r].
\end{equation}
%
%
Once the structure of polynomials $\hat{\Gamma}_i$, $q$, and $s_j,\,j=1,\ldots,n_c$, are fixed, \eqref{eq:SOSconditions} can in turn be expressed as a Linear Matrix Inequality (LMI) feasibility problem, where the decision variables are the coefficients of these polynomials, as proven in \cite{parrilo2000structured}. 

As previously mentioned, in this paper we are interested in computing level sets of RDLFs fully contained in $\mathcal{D}$ that are as large as possible. A possible way to do so is to maximize the integral of $\hat{\Gamma}_i(r)$, as
\begin{equation}\label{eq:Gammahat}
\left\{
\begin{array}{rl}
& \max\int\limits_{r\in\mathcal{R}_d}\hat{\Gamma}_i(r)\,\text{d}r\\
\text{s.t.}&\\
&V-\hat{\Gamma}_i+q\cdot c_i -\sum_{j=1}^{n_c}s_j\cdot\overline{c}_j\in\Sigma [x,r]\\
& \{s_j\}_{j=1}^{n_c}\in\Sigma [x,r]\\
& q \in\mathbb{R}[x,r]
\end{array}
\right..
\end{equation}
where $\mathcal{R}_d\subseteq\mathcal{R}$ is a compact domain, chosen so as to avoid improper integrals in the case where $\mathcal{R}$ is unbounded. Additionally, $\mathcal{R}_d$ can be chosen so as to prioritize the accuracy of $\hat{\Gamma}_i(r)$ in a certain region of interest of $\mathcal{R}$, \textit{e.g.}, around likely operation points of the system.

\begin{remark}
Note that whenever $\mathcal{R}_d$ is a normal domain \cite{calculus} described by polynomials, $\int_{r\in\mathcal{R}_d}\hat{\Gamma}_i(r)\,\textnormal{d}r$ can easily be computed in closed form and is polynomial \cite{algebra}, which implies that the objective function of \eqref{eq:Gammahat} remains linear in the coefficients of $\hat{\Gamma}_i$. When this is not the case, a practical approach is to randomly select a (possibly large) number of points in $\mathcal{R}_d$, $p_1,\ldots,p_{n_w}$, and use the following objective function
\begin{equation}\label{eq:objfun2}
    \frac{1}{n_w}\sum_{w=1}^{n_w} \hat{\Gamma}_i(p_w).
\end{equation}
Note that for a sufficiently large $n_w$, optimizing over \eqref{eq:objfun2} is equivalent to optimizing over $\int_{\mathcal{R}_d}\hat{\Gamma}_i(r)~\textnormal{d}r$. 
\end{remark}
\section{Example and practical considerations}\label{sec:practical}
It is widely known that the LMI problems that arise from SOS theory can become exceedingly sparse \cite{kojima2005sparsity} and numerically ill-conditioned. This may limit the applicability of the method presented in Section \ref{sec:method}. In this section we propose two techniques that might aid in mitigating numerical issues in the proposed methodology. The first technique is based on exploiting the structure of the problem at hand to utilize some \textit{a priori} knowledge in the structure of $\hat{\Gamma}(r)$. The second technique is based on a division of the domain $\mathcal{R}_d$.

In order to show the practical improvements  due to the application of these two techniques, we introduce a numerical case study consisting of a continuous time double integrator controlled with a PD control law 
\begin{equation}\label{eq:sys_example}
\dot{x}=\begin{bmatrix}
0 & 1\\
-\omega^2 & -2\zeta\omega
\end{bmatrix}x + \begin{bmatrix}0\\ \omega ^2
\end{bmatrix}r,
\end{equation}
where $\omega=10$ is the angular frequency and $\zeta=0.2$ is the damping ratio. This system is subject to the following constraint
\begin{equation}\label{eq:const_1}
c(x,r)=x_2-x_1^3+3x_1^2+10\geq0.
\end{equation}
Since $\overline{x}_r=[r\quad\!\!\!0]^T$, constraint \eqref{eq:const_1} defines $\mathcal{R}$ as $\mathcal{R}=\{r:r\leq3.721\}$. The domain $\mathcal{R}_d$ is chosen as $\mathcal{R}_d=\{r:-1.5\leq r \leq 3.721\}$. For every equilibrium point,  stability can be proved using a quadratic RDLF in the following form
\begin{equation}\label{eq:LF}
    V(x,r)=\left(x-\overline{x}_r\right)^\top \left[\begin{matrix}
12.6450&-0.005\\
-0.005&0.1263
\end{matrix}\right]\left(x-\overline{x}_r\right).
\end{equation}
\begin{figure}[t]
    \centering
    \includegraphics[width=\linewidth]{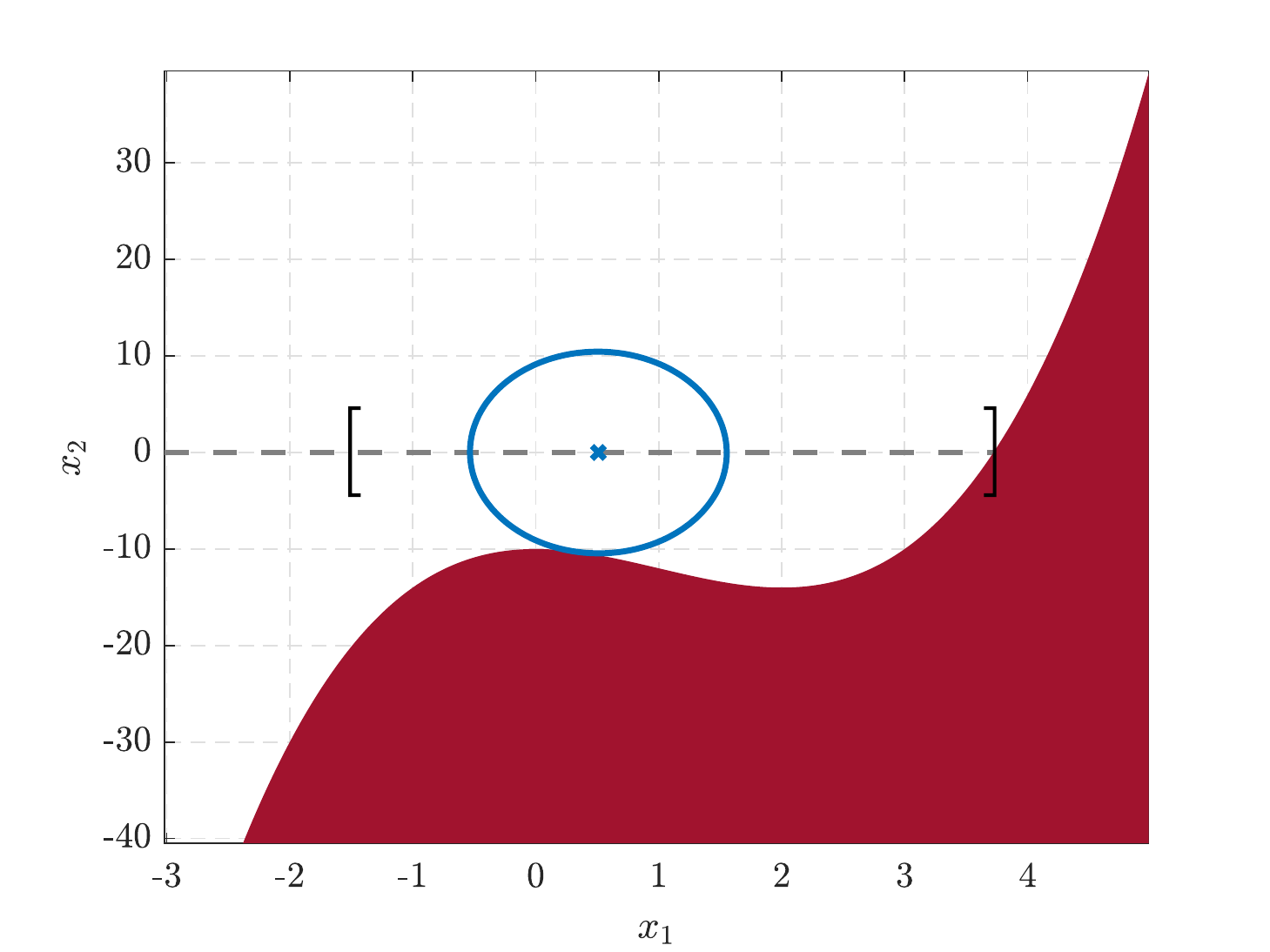}
    \caption{Visual representation of the case study. The set of possible points of equilibrium of system \eqref{eq:sys_example} is depicted as a grey dashed line, the set of points where $c(x,r)<0$ is represented by a solid dark red shaded area, and the set of points within $\mathcal{R}_d$ are encapsulated by black square brackets. For a given $r=\tilde{r}$, the set of points such that $V(x,\tilde{r})=\Gamma^\ast$ is represented as a solid blue ellipsoid, and the steady state associated to $\tilde{r}$, $\overline{x}_{\tilde{r}}$, is represented by a blue cross-shaped marker.}
    \label{fig:visual_representation}
\end{figure}
\begin{figure}[t]
    \centering
    \includegraphics[width=\linewidth]{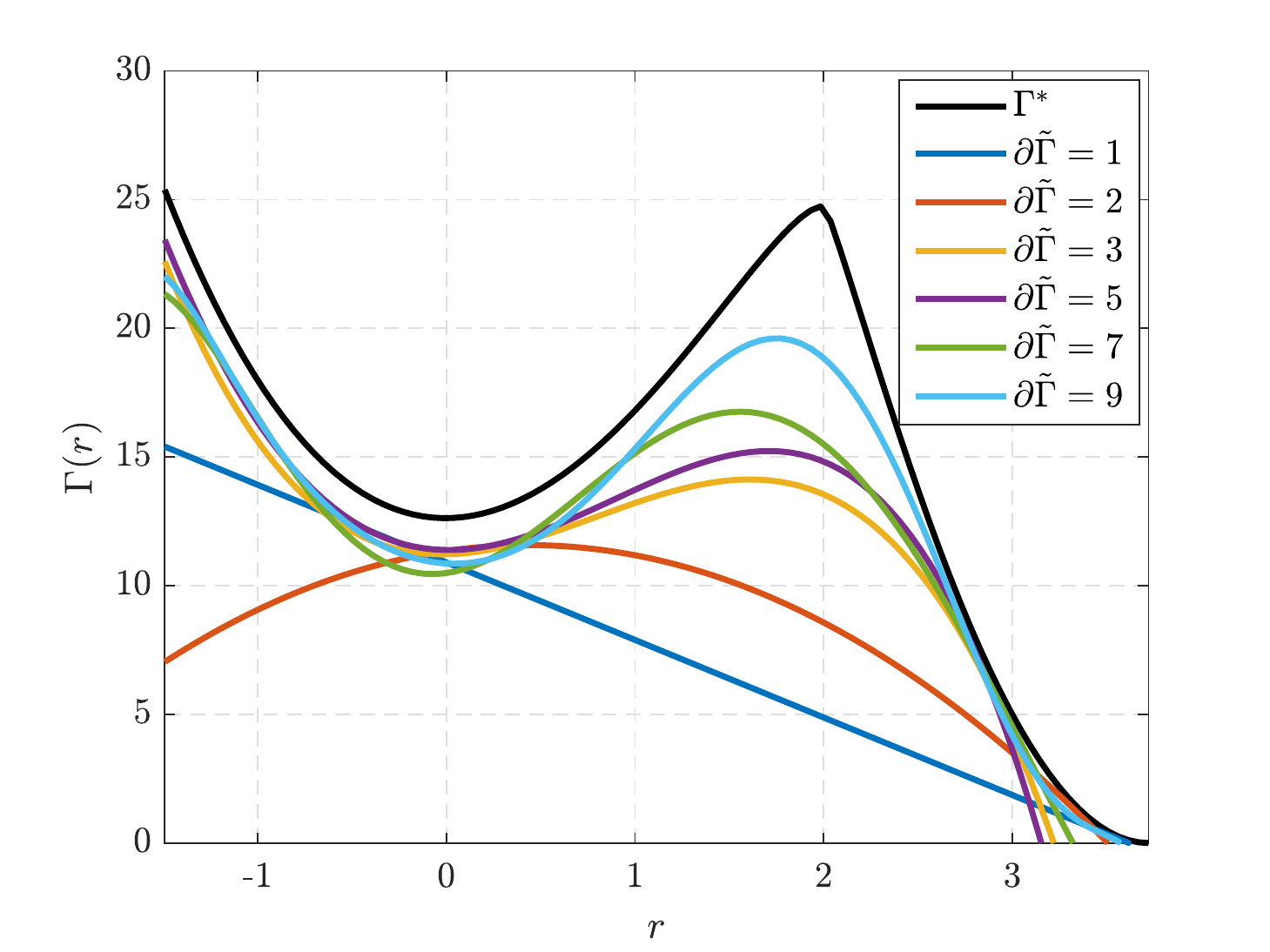}
    \caption{Approximation of $\Gamma^\ast$ for different values of $\partial\hat{\Gamma}$.}
    \label{fig:gamma_1poly_several_d}
\end{figure}
The problem is depicted in \figurename~\ref{fig:visual_representation}.  In \figurename~\ref{fig:gamma_1poly_several_d} we show the results obtained by solving \eqref{eq:Gammahat} for constraint \eqref{eq:const_1} for several  values of $\partial\hat{\Gamma}$. The obtained results are compared with $\Gamma^\ast(r)$ which was computed using a brute force approach. As expected, the accuracy of the approximation increases with the degree of $\hat{\Gamma}$. However, as it can be seen, a relatively high value of $\partial\hat{\Gamma}$ is required to get a fairly good approximation.
%
%
\subsection{Exploiting the structure of $\Gamma^\ast(r)$}
A first way to reduce the required degree of $\hat{\Gamma}_i(r)$ consists in exploiting the fact that the set $\{r: \Gamma^\ast_i(r)=0\}$ coincides with the set $\{r : \overline{c}_i(r)=0\}$. This is proved by the following lemma.
\begin{lemma}\label{lemma:signs}
Consider system \eqref{eq:system}, subject to the constraint $c_i(x,r)\geq0$, and whose global stability can be proved through an RDLF $V(x,r)$. Then for any $r\in\mathcal{R}$, $\overline{c}_i(r)>0$ implies that $\Gamma_i^\ast(r)>0$, and $\overline{c}_i(r)=0$ implies that $\Gamma^\ast_i(r)=0$.
\end{lemma}
\begin{pf}
Let $r'\in\mathcal{R}$ be such that $\overline{c}_i(r')>0$ (\textit{i.e.}, $r'$ is in the interior of $\mathcal{R}$) and $\overline{x}_{r'}$ its associated steady-state. The definition of $\Gamma^\ast_i(r)$ in \eqref{eq:optGammaStar} implies that
\begin{equation}\label{eq:lemma1}
\exists x' \in \{x:V(x,r')=\Gamma^\ast_i(r')\}  \textrm{, such that } c_i(x',r')=0.
\end{equation}
Since from the definition of RDLFs $V(x',r')>0$ if $x\neq\overline{x}_{r'}$, it follows from \eqref{eq:lemma1} that $\Gamma^\ast_i(r')>0$, and finally that $\overline{c}_i(r')>0\Leftrightarrow\Gamma^\ast_i(r')>0$. Using the same logic, it follows that $\Gamma^\ast_i(r'')=0\Leftrightarrow \overline{c}_i(r'')=0$ (\textit{i.e.}, $r''$ is on the border of $\mathcal{R}$). $\hfill\blacksquare$
\end{pf}
We can use the results of Lemma \ref{lemma:signs} to our advantage by assuming the following structure for $\hat{\Gamma}_i$
\begin{equation}\label{eq:cbar}
    \hat{\Gamma}_i=\overline{c}_i^k\cdot\tilde{\Gamma}_i,
\end{equation}
where $k\in\mathbb{Z}_{\geq0}$ and $\tilde{\Gamma}_i\in\mathbb{R}[r]$. Intuitively, the use of this term may require a lower $\partial\tilde{\Gamma}_i$ to achieve a higher accuracy, since assuming that $\hat{\Gamma}_i$ is factorized in $\overline{c}_i$ automatically sets $\hat{\Gamma}_i$ to 0 wherever $\overline{c}_i$ vanishes. 
\begin{figure}[t]
    \centering
    \includegraphics[width=\linewidth]{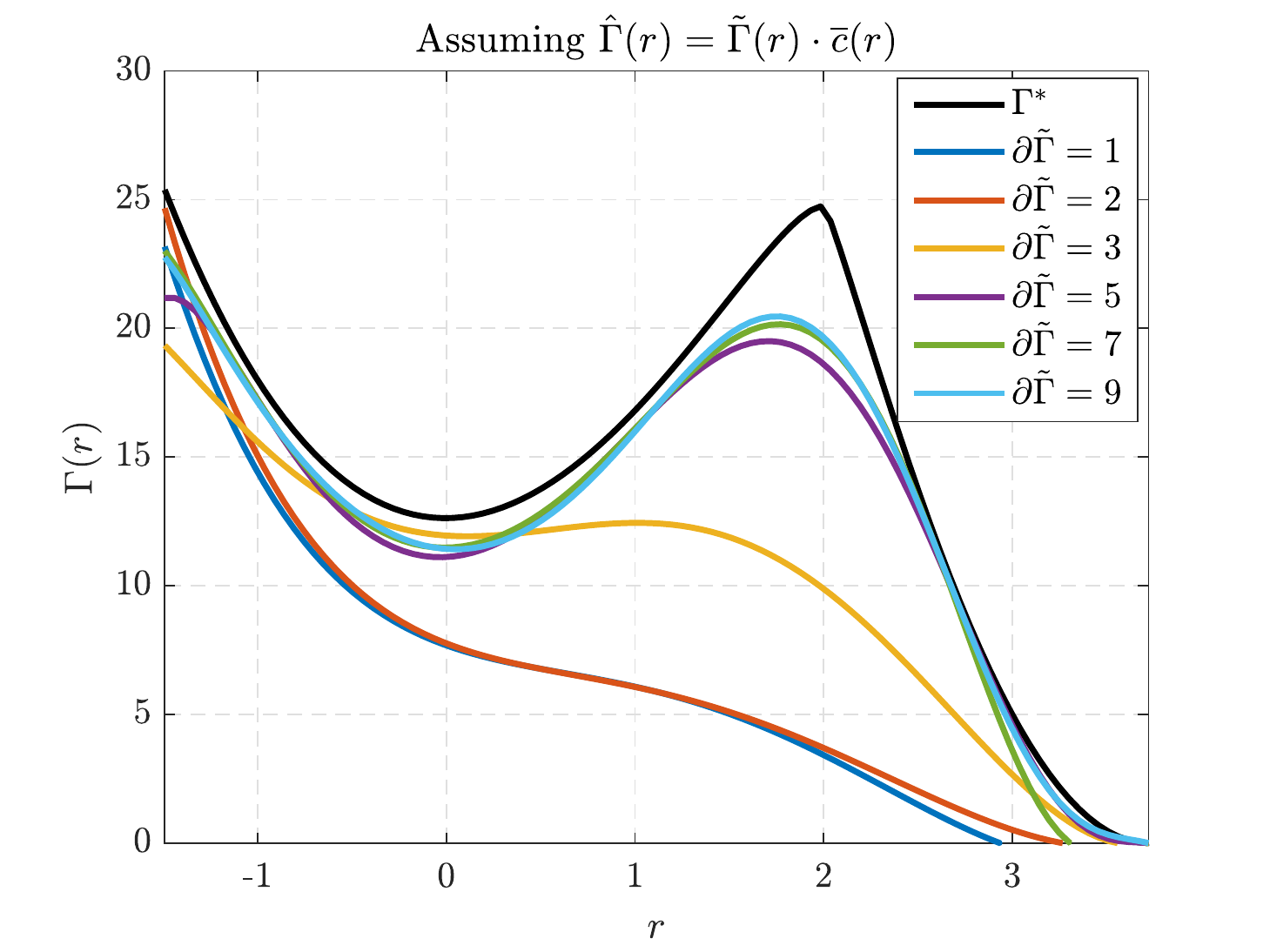}
    \caption{Approximation of $\Gamma^\ast$ for different values of $\partial\tilde{\Gamma}$, assuming $\hat\Gamma=\tilde{\Gamma}\cdot\overline{c}$.}
    \label{fig:gamma_1poly_several_d_cbar}
\end{figure}
\begin{figure}[t]
    \centering
    \includegraphics[width=\linewidth]{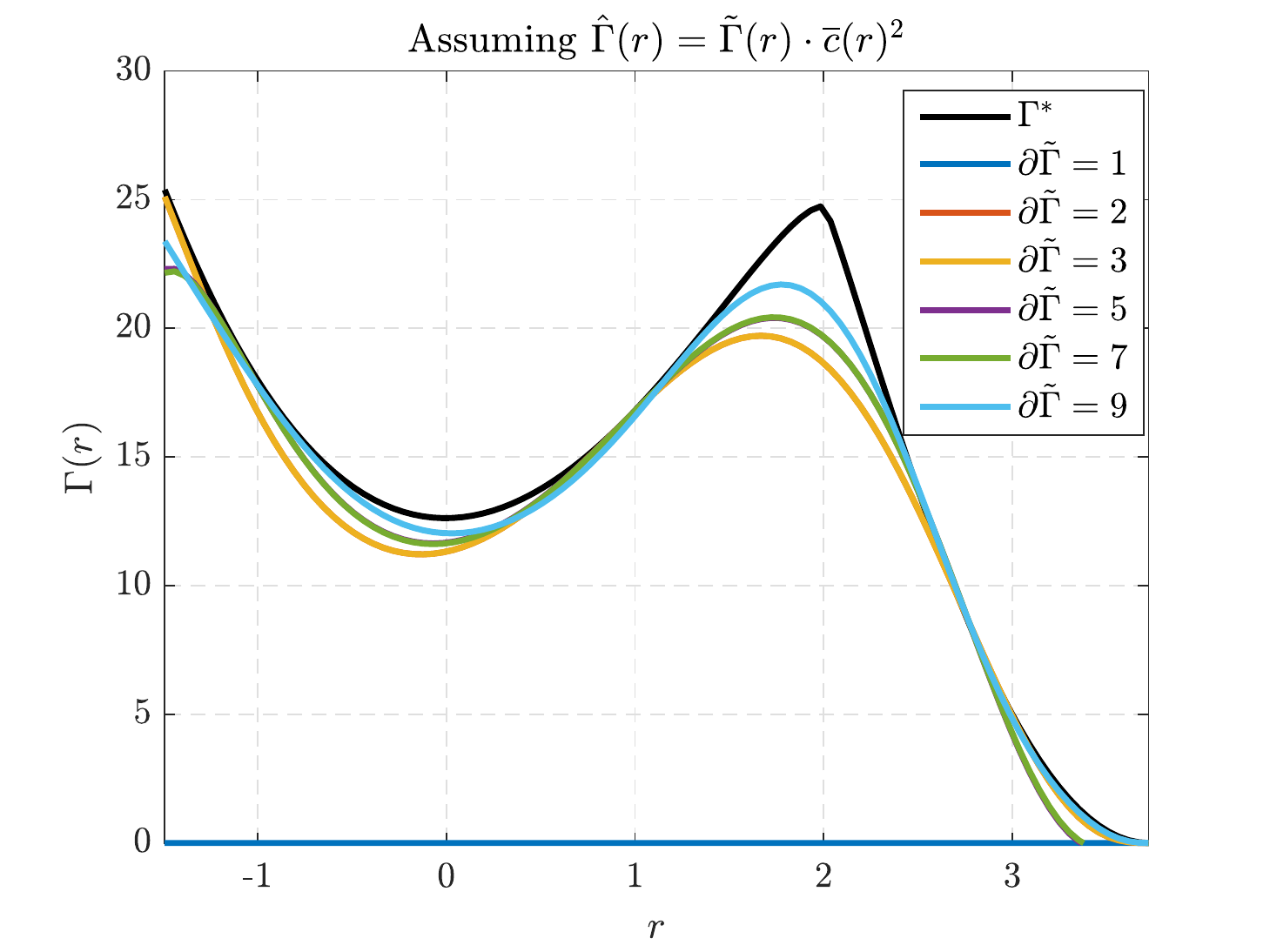}
    \caption{Approximation of $\Gamma^\ast$ for different values of $\partial\tilde{\Gamma}$, assuming $\hat\Gamma=\tilde{\Gamma}\cdot\overline{c}^2$. Note that the curve corresponding to $\partial\tilde{\Gamma}=2$ overlaps with $\partial\tilde{\Gamma}=3$, and $\partial\tilde{\Gamma}=5$ overlaps with $\partial\tilde{\Gamma}=7$.}
    \label{fig:gamma_1poly_several_d_cbar2}
\end{figure}
\begin{figure}[t]
    \centering
    \includegraphics[width=\linewidth]{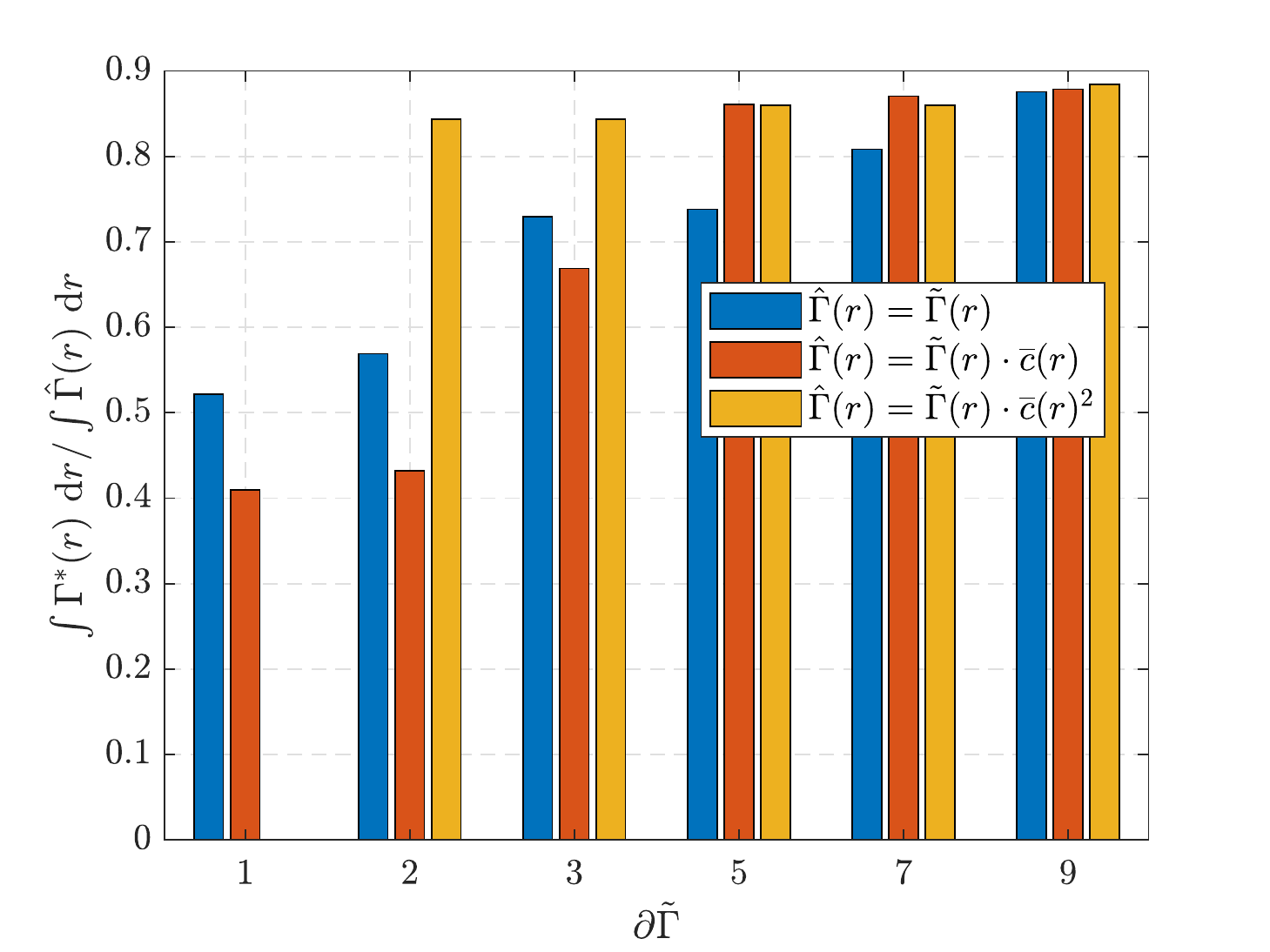}
    \caption{Accuracy of $\hat{\Gamma}=\overline{c}^k\cdot\tilde{\Gamma}$ with respect to the degree of $\tilde{\Gamma}$ for $k\in\{0,1,2\}$.}
    \label{fig:bargraph}
\end{figure}
In \figurename~\ref{fig:gamma_1poly_several_d_cbar} and \ref{fig:gamma_1poly_several_d_cbar2} we show the results of applying this notion to our case study, assuming that $\hat{\Gamma}$ is factorized in $\overline{c}$ and $\overline{c}^2$, respectively. As it can be seen in \figurename~\ref{fig:bargraph}, when assuming $\hat{\Gamma}=\tilde{\Gamma}\cdot\overline{c}$ the required degree to obtain an accuracy of more than 80\% is $\partial\tilde{\Gamma}=5$, compared to the required degree of $\partial\hat{\Gamma}=7$ if this assumption is not made. The required degree is further decreased to $\partial\tilde{\Gamma}=2$ if we assume $\hat{\Gamma}=\tilde{\Gamma}\cdot\overline{c}^2$.
\subsection{Piece-wise polynomial approach}\label{sec:practical_div}
In some cases the Semi-Definite Programming (SDP) solver may not arrive at a sensible approximation of $\Gamma^\ast_i$ due to the high dimensionality of the system, to the high degree of the involved polynomials, or to the resulting optimization problem being ill-conditioned. To tackle this, a possible approach is to divide $\mathcal{R}_d$ in several subsets $\mathcal{R}_\ell,~\ell=1,\ldots,n_r$ such that
\begin{equation}
\mathcal{R}_d=\bigcup_{\ell=1}^{n_r} \mathcal{R}_\ell. 
\end{equation}
These subsets are described by the additional constraints $\overline{c}^\ell_j(r)$, with $j=1,\ldots,n_\ell$. Subsequently the polynomial approximation of $\Gamma^\ast$ associated with the $\ell$-th subset $\mathcal{R}_\ell$, denoted by $\hat{\Gamma}_{i,\ell}$, can be computed through the following SOS condition
\begin{equation}\label{eq:gamma_Rd}
        V-\hat{\Gamma}_{i,\ell}+q\cdot c_i - \sum_{j=1}^{n_\ell} s_j \cdot \overline{c}_j \in \Sigma [x,r],
\end{equation}
Once the $\hat{\Gamma}_{i,\ell}$s have been computed, $\hat{\Gamma}_i(r)$ can be recovered as follows:
\begin{equation*}
    \hat{\Gamma}_i(r)=\max_\ell\{\hat{\Gamma}_{i,\ell}(r)\},
\end{equation*}
where $\hat{\Gamma}_{i,\ell}=0~\forall r \notin \mathcal{R}_\ell$. By dividing $\mathcal{R}_d$, not only we reduce the size of the domain over which we compute each $\hat{\Gamma}_{i,\ell}$, but also the term $\sum_{j=1}^{n_\ell} s_j \cdot \overline{c}_j$ becomes simpler since typically $n_c>n_\ell$.
A possible approach for the division of $\mathcal{R}_d$ is the Delaunay tessellation \cite{watson1981computing,lawson1986properties}, which divides the space into $p$-simplices. It is well known that a $p$-simplex requires $p+1$ inequalities to be described, therefore, this tessellation achieves the minimal amount of inequalities describing a given inner $\mathcal{R}_\ell$ for a polyhedral division of $\mathcal{R}_d$.

\begin{figure}[t]
    \centering
    \includegraphics[width=\linewidth]{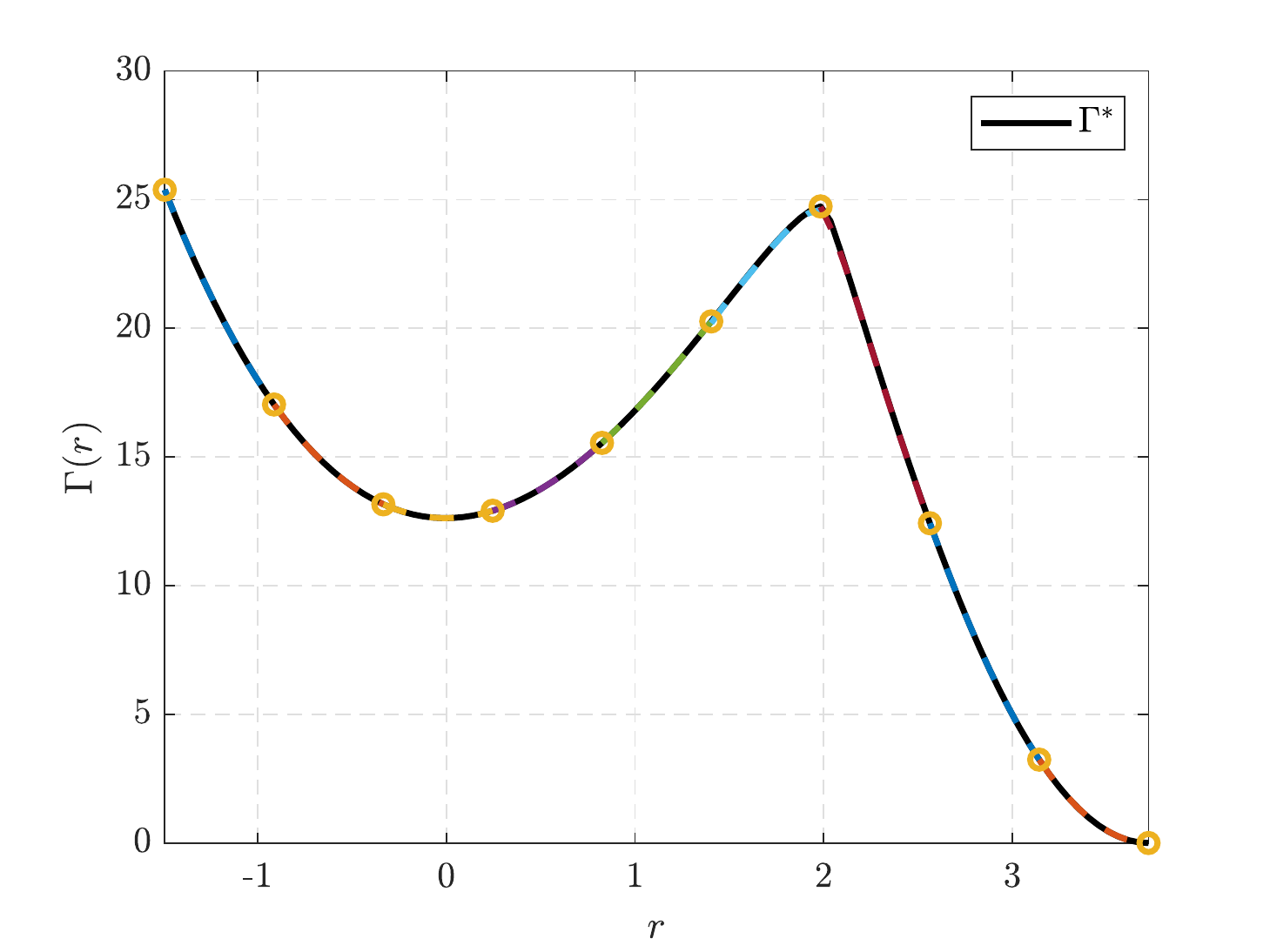}
    \caption{Piece-wise polynomial approach for the case study, using $\partial\hat{\Gamma}_\ell=4,~ \ell\in \{1,\ldots,9\}$. $\Gamma^\ast$ is depicted as a solid black line, each of the divisions is marked as a yellow circular marker, and every $\hat{\Gamma}_\ell$ is depicted as a dashed colored line.}
    \label{fig:several}
\end{figure}

We show the results of applying this approach to the case study in \figurename~\ref{fig:several}, for $n_r=9$, $\partial\hat{\Gamma}_{1,\ell}=\partial s_j=4$. As it can be seen, $\hat{\Gamma}$ and $\Gamma^\ast$ overlap, thus yielding a very good approximation for this case study.
\begin{remark}
It should be noted that the two techniques presented in this section are not mutually exclusive and can be used at the same time.
\end{remark}
\section{Applications in constrained control}\label{sec:app}
In this section we will discuss two different applications of the proposed reference dependent positively invariant sets in constrained control problems. The first application is in discrete time and allows us to compute the terminal set in the MPC for Tracking framework for discrete time systems. The second case shows that continuous time positively invariant sets can be used to calculate the dynamic safety margin in the ERG framework.  
\subsection{Safe Invariant Sets As Terminal Conditions In the MPC for Tracking Framework}\label{sec:appMPC}
In the last two decades, MPC schemes have been widely used to address constrained control problems \cite{Mayne2000,Camacho2007}. These schemes have the remarkable feature of being able to drive the system state to a fixed set-point while optimizing the control performances and satisfying the constraints of the system.

However, classical MPC schemes might lose feasibility under sudden set-point changes \cite{bemporad1997nonlinear}. To tackle this, the MPC for Tracking scheme was presented in \cite{limon2008mpc}. This control scheme solves at every time step $t$ the following optimization problem
\begin{IEEEeqnarray}{lCr}
\min_{\textbf{u},v}\  J_{N_p,N_c}(x,r;\textbf{u},v), \IEEEyesnumber \IEEEyessubnumber \label{eq:MPC_first}\\
\textrm{s.t.}\nonumber \\
x(0)=x(t),\IEEEnonumber \IEEEyessubnumber\\
x(j+1)=\phi(x(j),u(j)),\quad j=0,\ldots,N_c-1, \IEEEnonumber \IEEEyessubnumber\\
(x(j),u(j))\in\mathcal{Z},\quad j=0,\ldots,N_c-1, \IEEEnonumber \IEEEyessubnumber\\
x(j+1)=\phi(x(j),\kappa(x(j),v)),\, j=N_c,\ldots,N_p-1, \IEEEnonumber \IEEEyessubnumber \IEEEeqnarraynumspace\\
(x(j),\kappa(x(j),v))\in\mathcal{Z},\quad j=N_c,\ldots,N_p-1, \IEEEnonumber \IEEEyessubnumber\\
\overline{x}_v=g_x(v),\IEEEnonumber \IEEEyessubnumber\\
\overline{u}_v=g_u(v), \IEEEnonumber \IEEEyessubnumber\\
(x(N_p),v)\in\Omega, \IEEEnonumber \IEEEyessubnumber\label{eq:MPC_last}
\end{IEEEeqnarray} 
with objective function
\begin{IEEEeqnarray}{lCr}
J_{N_p,N_c}(x,r;\textbf{u},v)=\sum_{j=0}^{N_c-1}J_s(x(j)-\overline{x}_v,u(j)-\overline{u}_v)\nonumber\\ 
+\sum_{j=N_c}^{N_p-1}J_s(x(j)-\overline{x}_v,\kappa(x(j),v)-\overline{u}_v) \\
+J_f(x(N_p)-\overline{x}_v)+J_o(v-r) \nonumber
\end{IEEEeqnarray}
where $N_p$ and $N_c$ are the prediction and control horizons, respectively, $\kappa:\mathbb{R}^n\times\mathbb{R}^p\rightarrow\mathbb{R}^m$ is a stabilizing control law as in Section \ref{sec:PS}, $\mathcal{Z}=\mathcal{X}\times\mathcal{U}$, $v\in\mathcal{R}$ is the auxiliary reference, $g_x:\mathcal{R}\rightarrow\mathbb{R}^n$ and  $g_u:\mathcal{R}\rightarrow\mathbb{R}^m$ are two locally Lipschitz functions that map the auxiliary reference to its corresponding steady state and input $(\overline{x}_v,\overline{u}_v)$ (\textit{i.e.}, $\overline{x}_v=\phi(g_x(v),g_u(v)$),  $J_s:\mathbb{R}^n\times\mathbb{R}^m\rightarrow\mathbb{R}$,  $J_o:\mathbb{R}^p\rightarrow\mathbb{R}$, and $J_f:\mathbb{R}^n\rightarrow\mathbb{R}$ are convex positive definite functions that represent the stage, offset, and terminal costs, respectively (see \cite{limon2018nonlinear} for more details). $\Omega$ is an invariant set for tracking which can be defined as follows:
\begin{definition}\label{def:invset4t}
For a given set of constraints $\mathcal{Z}=\mathcal{X}\times\mathcal{U}$, a set of admissible references $\mathcal{R}$, and a local control law $u=\kappa(x,v)$, a set $\Omega\subset\mathbb{R}^n\times\mathbb{R}^p$ is an (admissible) invariant set for tracking for system \eqref{eq:system} if for all $(x,v)\in\Omega$, we have  $(x,\kappa(x,v))\in\mathcal{Z}$, $v\in\mathcal{R}\ominus\mathcal{B}_p(\varepsilon)$, and $(\phi(x,\kappa(x,v)),v)\in\Omega$.
\end{definition}
This set represents the MPC for Tracking counterpart of the classical terminal set used in MPC schemes to guarantee stability and recursive feasibility. The computation of such a reference dependent set is one of the most challenging problems when designing an MPC for tracking. 

To approximate these sets, in \cite{limon2018nonlinear} the authors resort to a partition of the set of points of equilibrium paired with linearization of the system for each partition based on a Linear Time Varying characterization \cite{wan2003efficient,wan2003efficient2}. However, this \textit{ad-hoc} technique is conservative and has a solution only for limited classes of constraints. 

\begin{figure}
    \centering
    \includegraphics[width=\linewidth]{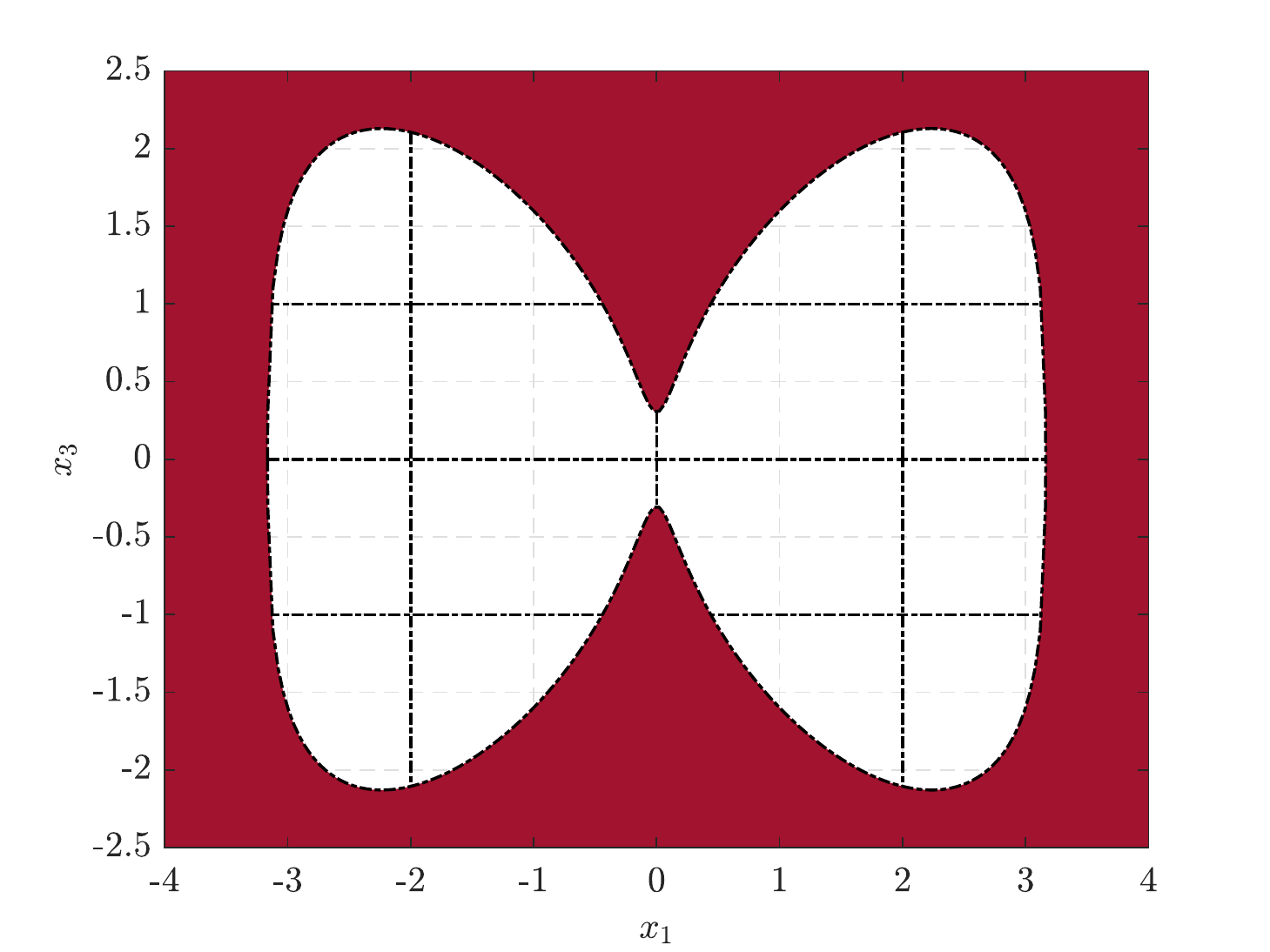}
    \caption{Representation of $\mathcal{R}$ and its division used in the example at hand.}
    \label{fig:domain_mpc}
\end{figure}

Note that once a control law $\kappa(x,v)$ acting as a terminal control law has been fixed \cite{Schmid2010,Ntogramatzidis2016}, terminal sets for this scheme can be characterized as safe level sets of RDLFs, meaning that the proposed method in Section \ref{sec:method} can be employed to compute these sets.  In the rest of this subsection we will show an example where an invariant set for tracking is computed using the methodology proposed in this paper.
\paragraph*{Example:}
Consider the following model of a ball-and-plate system in discrete time
\begin{align}\label{eq:sys_sim_MPC}
x(t+1)=&\left[\begin{matrix}
1 & 0.5 & 0 & 0\\
0 & 1 & 0 & 0\\
0 & 0 & 1 & 0.5\\
0 & 0 & 0 & 1
\end{matrix}\right]x(t)+\left[\begin{matrix}
0.125 & 0\\
0.5 & 0\\
0 & 0.125\\
0 & 0.5
\end{matrix}\right]u(t).
\end{align}
The positions of this system are constrained as follows
\begin{equation}\label{eq:mpc_const}
c(x,v)=x_1^4+x_3^4-10x_1^2+x_3^2-0.1\leq0,
\end{equation}
which describes the \textit{bow tie} set depicted in \figurename~\ref{fig:domain_mpc}. In order to build the MPC for Tracking, we chose the following stabilizing terminal control law
\begin{equation}
\kappa(x,v)=\begin{bmatrix}
-4 & -2.73 & 0 & 0\\
0 & 0 & -4 & -2.73
\end{bmatrix}
x+\begin{bmatrix}
4 & 0\\
0 & 4
\end{bmatrix}v.
\end{equation}
When the system is controlled with this law and $v$ is kept constant, for every equilibrium point $\overline{x}_v=[v_1\ 0\ v_2 \ 0]^T$, it admits the following Lyapunov function
\begin{equation}\label{eq:LyapunovExample}
V(x,v)=(x-\overline{x}_{v})^TP(x-\overline{x}_{v}),
\end{equation}
with
\begin{equation}
P=\left[\begin{matrix}
    5.3933 & 0.8668 & 0 & 0\\
    0.8668 & 1.1946 & 0 & 0\\
    0 & 0 & 5.3933 & 0.8668\\
    0 & 0 & 0.8668 & 1.1946
\end{matrix}\right].
\end{equation}
To compute $\hat{\Gamma}$, we used the approach presented in Section \ref{sec:practical_div}: we divided $\mathcal{R}$ into 16 subsets (see \figurename \ref{fig:domain_mpc}), and we did not assume any parameterization on $\overline{c}$ (\textit{i.e.}, $k=0$ in \eqref{eq:cbar}). For what concerns the maximal degrees of the decision polynomials, we set $\partial q_\ell=6$, $\partial s_{j,
\ell}=4$ and $\partial \hat{\Gamma}_\ell=8$, where $\ell\in\{1,\ldots,16\}$.

Since the set described by \eqref{eq:mpc_const} is non-convex, we used the results in \cite{cotorruelo2018tracking} to ensure convergence. For what concerns the MPC for Tracking design parameters, the control and prediction horizons were set to $N_c=1$ and $N_p=2$, respectively. The stage and offset cost functions in the objective function of the MPC optimization problem are
\begin{IEEEeqnarray*}{lCr}
J_s(x-\overline{x}_v,u-\overline{u}_v)=(x-\overline{x}_v)^TQ(x-\overline{x}_v)+\\(u-\overline{u}_v)^TR(u-\overline{u}_v),\\
J_o(v-r)=(v-r)^TT(v-r),\\
J_f(x-\overline{x}_v)=(x(N_p)-\overline{x}_v)^TP(x(N_p)-\overline{x}_v),
\end{IEEEeqnarray*}
with weights $Q=I_4$, $R=0.1I_2$, and $T=10I_2$.

\begin{figure}[!t]
    \centering
    \includegraphics[width=\linewidth]{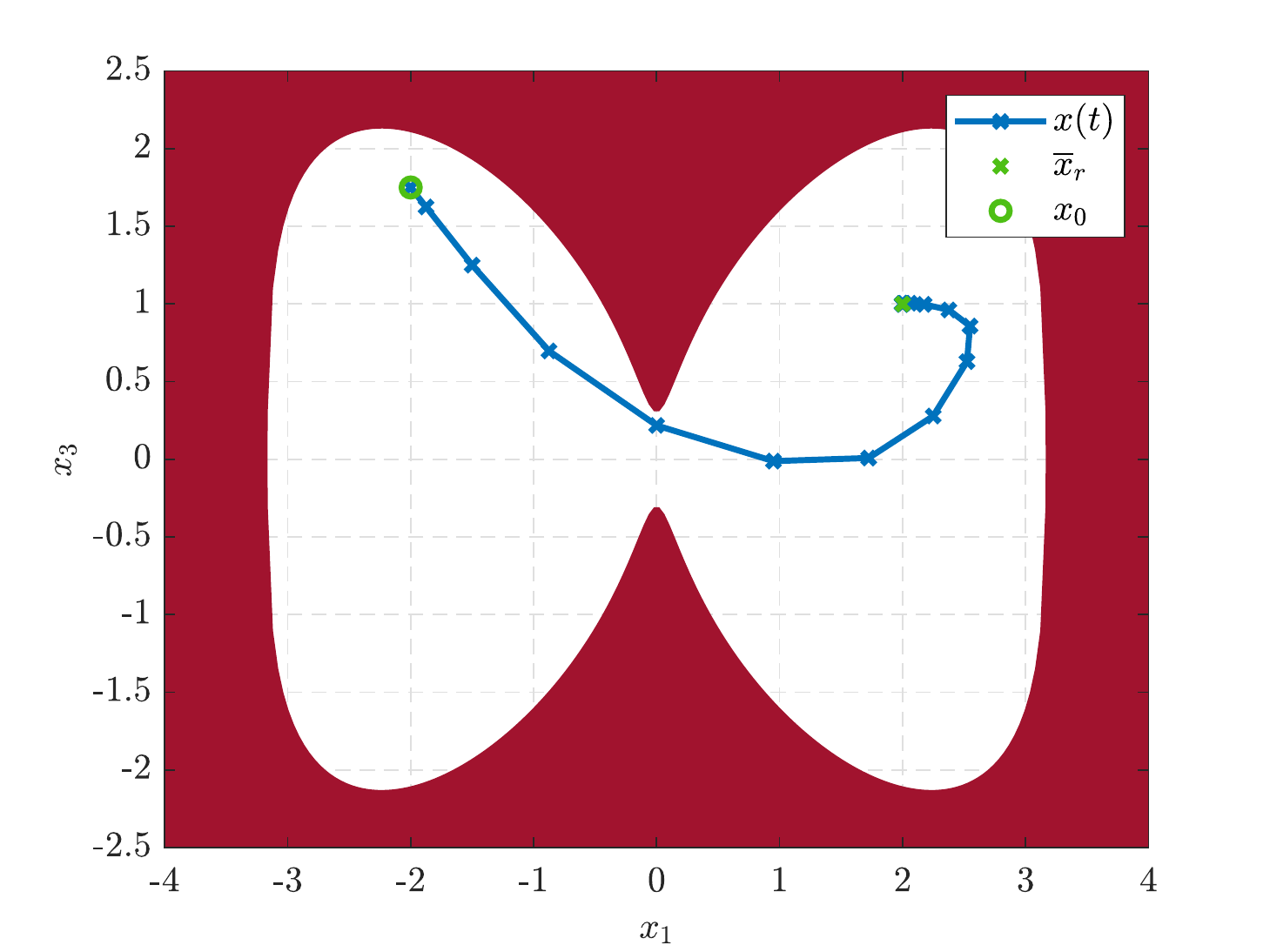}
    \caption{Simulation results; state trajectory of the system when controlled with a MPC for Tracking scheme. The state trajectory is depicted as a solid blue line with cross markers. The initial and steady state associated to the desired reference are depicted in green circular and cross-shaped markers, respectively.}
    \label{fig:mpc_phase_plot}
\end{figure}

\begin{figure}[!t]
    \centering
    \includegraphics[width=\linewidth]{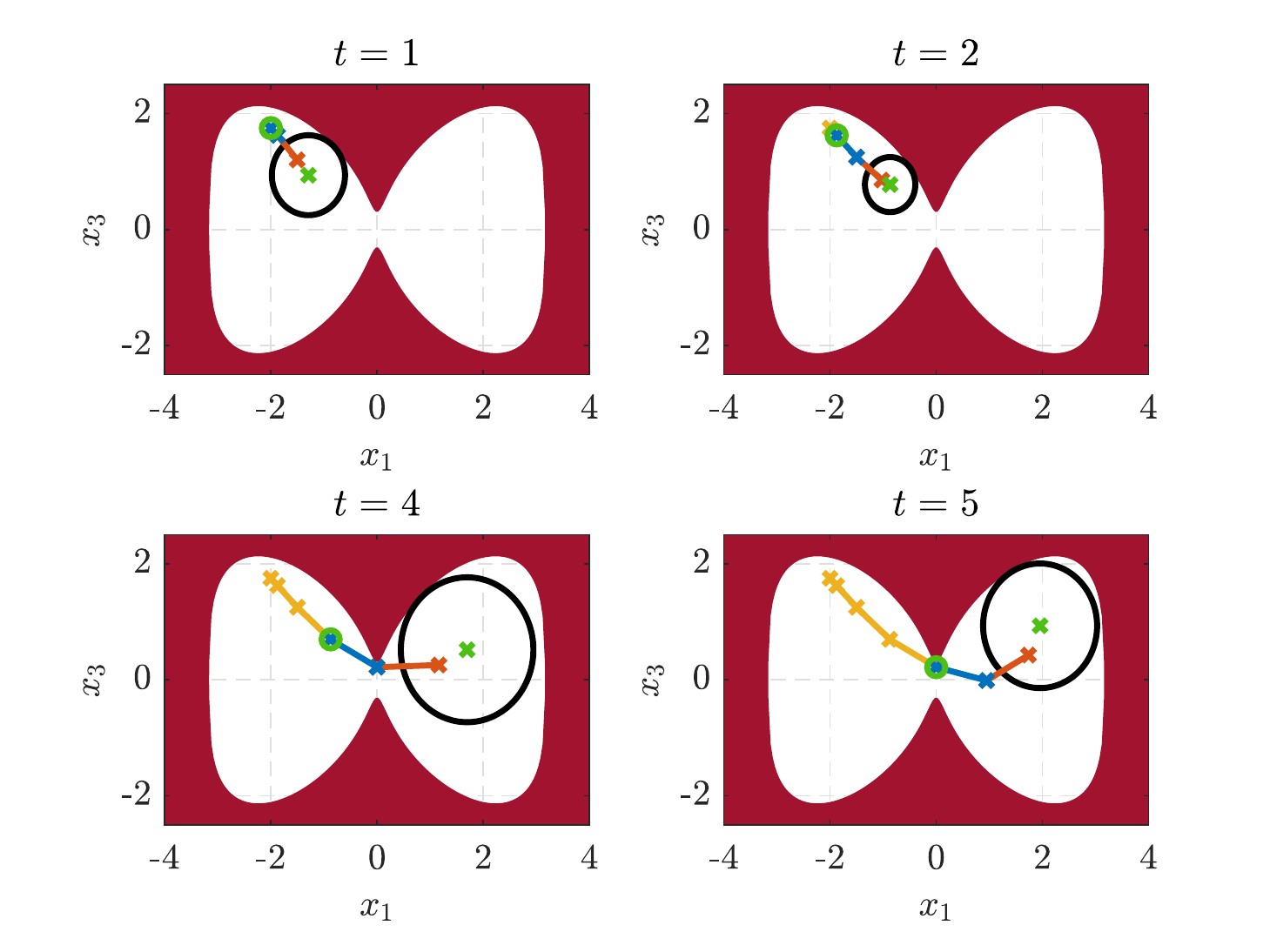}
    \caption{Simulation results; previous state trajectory, its predicted evolution and terminal set at different time steps. The previous state trajectory is depicted in a solid yellow line, the predicted trajectory for the first $N_c$ steps is depicted in a solid blue line, the last $N_p-N_c$ steps is depicted in a solid orange line, and the terminal invariant set is outlined in a solid black line. The initial point is marked with a green circle, and the steady state associated to the auxiliary reference is represented with a green cross.}
    \label{fig:subplots}
\end{figure}

In \figurename~\ref{fig:mpc_phase_plot} and \ref{fig:subplots} we report the simulation results of the resulting MPC scheme assuming the initial state $x_0=[-2\ 0\ 1.75 \ 0]^T$, and setting the reference to $r=[2\ 1]^T$. As it can be seen in \figurename~\ref{fig:subplots} (which shows only for some time steps) the MPC is able to drive the system in $N_p$ steps to the terminal invariant set, whose size allows for a large domain of attraction of the MPC for Tracking.
\subsection{Reference Dependent Positively Invariant Sets in the ERG Framework}\label{sec:appERG}
\begin{figure}[!t]
    \centering
    \includegraphics[width=\linewidth]{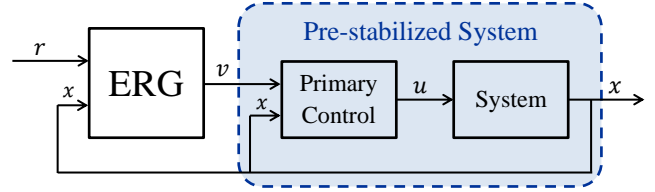}
    \caption{The general structure of the ERG scheme.}
    \label{fig:ERG}
\end{figure}

The ERG \cite{Nicotra2018} is an add-on unit that is able to provide constraint satisfying capabilities to prestabilized continuous time systems (see \figurename~\ref{fig:ERG}). The main idea behind the ERG is to feed the precompensated system with a filtered version of the desired reference $r$, $v\in\mathbb{R}^p$, computed such that if $v$ remains constant, the system will not violate any constraints. In particular, the ERG \cite{Hosseinzadeh2018} manipulates the time derivative of the auxiliary reference $v$ as
\begin{equation}\label{eq:ERG}
\dot{v}=\Delta(x,v)\cdot\rho(r,v),
\end{equation}
where $\rho(r,v)$ and $\Delta(x,v)$ are the two fundamental components of the ERG, called the Navigation Field (NF) and the Dynamic Safety Margin (DSM), respectively.

The NF is a vector field such that for any two steady-state admissible references $v$ and $r$, the trajectory of system $\dot{v}=\rho(r,v)$ goes from $v$ to $r$ through a path of strictly steady-state admissible references. This problem can be addressed using standard path planning algorithms (\textit{e.g.}, \cite{latombe2012robot}).

The DSM is a measure of the distance between the constraints and the system trajectory that would emanate from the state $x$ for a constant reference $v$. A possible way to construct a DSM \cite{Nicotra2018,Hosseinzadeh2018} is by using an RDLF $V(x,v)$ and a bound $\hat{\Gamma}$ such that $\mathcal{S}_{\hat{\Gamma}}$ is a safe invariant set as follows:
\begin{equation}\label{DSM_formula}
\Delta(x,v)=\lambda\cdot(\hat{\Gamma}(v)-V(x,v)),
\end{equation}
where $\lambda>0$ is a tuning parameter. Note that this implies that $\Delta(x,v)>0$ whenever $x$ is in the interior of $\mathcal{S}_{\hat{\Gamma}}(v)$, and $\Delta(x,v)=0$ when $x$ is on the border of $\mathcal{S}_{\hat{\Gamma}}(v)$.

Accordingly, the methodology presented in this paper can be applied directly to compute the bound $\hat{\Gamma}(v)$ in the DSM within the ERG framework.
\paragraph*{Example:}
Consider a ball-and-plate system in continuous time stabilized with a PD control law
\begin{equation}
\dot{x}=\begin{bmatrix}
     0 & 1 & 0 & 0\\
  -100 & -4 & 0 & 0 \\
     0 & 0 & 0 & 1\\
     0 & 0 & -100 & -4
\end{bmatrix}x + \begin{bmatrix}
0 & 0\\
100 & 0\\
0 & 0\\
0 & 100
\end{bmatrix}v,
\end{equation}
whose positions are constrained to lie within the \textit{bow tie} set \eqref{eq:mpc_const}. For every equilibrium point $\overline{x}_v=[v_1\ 0\ v_2 \ 0]^T$, stability can be proved using the quadratic RDLF
\begin{equation*}
V(x,v)=(x-\overline{x}_v)^TP(x-\overline{x}_v),
\end{equation*}
with
\begin{equation*}
P=\left[\begin{matrix}
12.645 & 0.005 & 0 & 0\\
0.005 & 0.1263 & 0 & 0\\
0 & 0 & 12.645 & 0.005\\
0 & 0 & 0.005 & 0.1263
\end{matrix}\right].
\end{equation*} 
We computed $\hat{\Gamma}$ by dividing $\mathcal{R}$ and by setting the degrees of the polynomials in the same manner as in the example presented in Section \ref{sec:appMPC}.

\begin{figure}[!t]
\centering
\includegraphics[width=\linewidth]{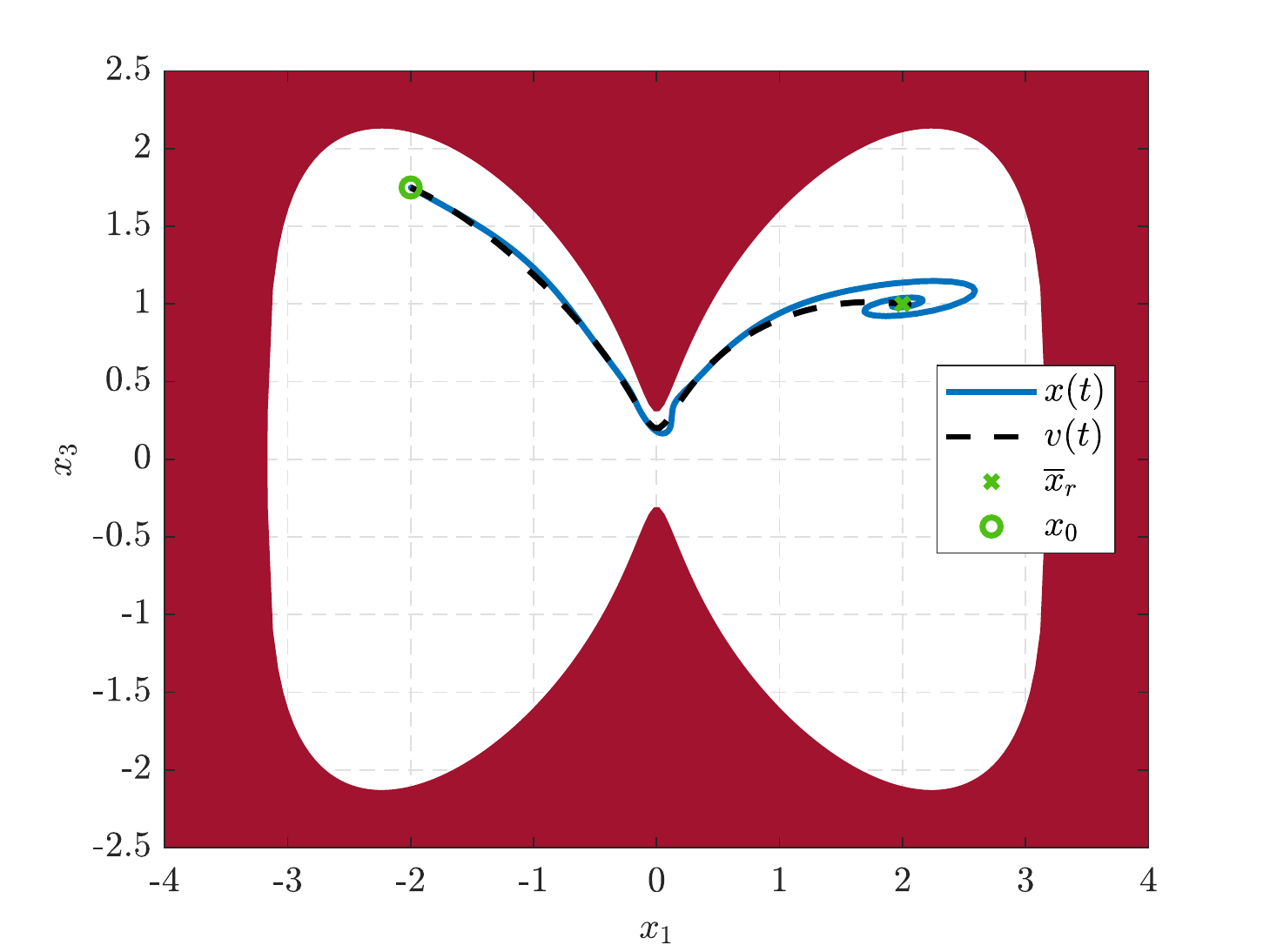}
\caption{Simulation results; state trajectory of the system when controlled with an ERG. The trajectory of the system is depicted in solid blue line, the trajectory of the auxiliary reference,$v$ , is depicted as a black dashed line. The initial and steady state associated to the desired reference are depicted in green circular and cross-shaped markers, respectively.}\label{fig:result1}
\end{figure}

For what concerns the NF, since $\mathcal{R}$ is non-convex\cite{Nicotra2018}, it is enough to choose it as
\begin{equation*}
\rho(r,v)=\frac{\nabla\Phi(v)^{-1}(\Phi(r)-\Phi(v))}{\max\{\nabla\Phi(v)^{-1}(\Phi(r)-\Phi(v)),\theta\}},
\end{equation*}
where  $\Phi:\mathcal{R}_d\rightarrow\mathcal{R}_c$ is a diffeomorphism that maps the interior of $\mathcal{R}$ to a convex set $\mathcal{R}_c$ and $\theta=0.01$ is a smoothing factor. A possible choice for $\Phi$ is
\begin{equation*}
\Phi(r)= \begin{bmatrix} r_1\\\displaystyle\frac{r_2}{\sqrt{\displaystyle\frac{1}{2}\sqrt{- 4r_1^4 + 40r_1^2 + \displaystyle\frac{7}{5}} - \displaystyle\frac{1}{2}}}
\end{bmatrix}.
\end{equation*}
Simulation results are shown in \figurename~\ref{fig:result1} for an initial state $x_0=[-2\ 0\ 1.75 \ 0]^T$, and a desired reference $r=[2\ 1]^T$. As it can be seen, the proposed ERG is able to steer the system state to the desired reference point while fulfilling the constraints at all times. The reader is referred to \url{https://www.youtube.com/watch?v=25yPHpviR18} for a video containing extra material. 
\section{Conclusions}\label{sec:conclusion}
In this paper we proposed a systematic method to approximate safe reference dependent positively invariant sets parameterized for systems that are subject to general polynomial constraints.

To do so, first, we demonstrated that such sets can be determined through an optimization problem. It was then shown that we can approximate a parameterized solution of this optimization problem by making use of SOS techniques. We later showed that it is possible to alleviate some of the numerical issues that the SOS framework presents if the underlying structure of the largest safe level set is exploited, and if the overall optimization problem is broken down into several, smaller better numerically conditioned problems. 

The proposed method has relevant applications in constrained control schemes. In particular, the proposed method can be used in the MPC for Tracking framework to determine the terminal set, and in the ERG framework to determine the Dynamic Safety Margin. Corresponding formulations in the context of the mentioned applications were discussed, and a numerical simulation for each of the mentioned applications were presented in order to evaluate the effectiveness of the proposed method.

We believe that the proposed methodology can be applied to systems subject to polynomial constraints such as robotic manipulators or the aging-aware charge of Li-ion batteries \cite{romagnoli2017computationally}. Possible future research lines include  simultaneously computing the RDLF and its largest safe level set and generalizing the methodology beyond polynomials, \textit{e.g.} for rational functions. 
\bibliography{ref}
\appendix
\section{Krivine--Stengle Positivstellensatz}\label{sec:psatz}
Before presenting the Krivine--Stengle Positivstellensatz, a few definitions need to be introduced. For the sake of simplicity, the following concepts will not be explained in depth, and will rather be mathematically characterized. For further information on the matter, the reader is referred to \cite{parrilo2000structured,bochnak2013real,stengle1974nullstellensatz} and the references therein.

\begin{definition}{(Multiplicative Monoid)}
The multiplicative monoid generated by a set of polynomials $P=\{p_i\}_{i=1}^m$, $p_i\in\mathbb{R}[x_1,\ldots,x_n],~ i=1,\ldots,m$ is the set of finite products of the $p_i$, including the unity:
\begin{equation*}
    \textnormal{Monoid}(P)=\left\{\prod_{i=1}^m p_i^{k_i},\  k_i\in\mathbb{Z}_{\geq 0},~ i=1,\ldots,m\right\}.
\end{equation*}
\end{definition}
\begin{definition}{(Cone)}
The cone generated by a set of polynomials $P=\{p_i\}_{i=1}^m$ , $p_i\in\mathbb{R}[x_1,\ldots,x_n],~i=1,\ldots,m$ is the sum of the elements of $\textnormal{Monoid}(P)$ multiplied by some sum of squares polynomials $s_i$:
\begin{multline*}
    \textnormal{Cone}(P)=\left\{ s_0 + \sum_i s_i g_i : s_i \in \Sigma[x_1,\ldots,x_n], \right. \\ g_i \in \textnormal{Monoid}(P) \Bigg\}.
\end{multline*}
\end{definition}
\begin{definition}{(Ideal)}
The ideal generated by a set of polynomials $P=\{p_i\}_{i=1}^m$, $p_i\in\mathbb{R}[x_1,\ldots,x_n],~i=1,\ldots,m$ is the sum of the products of the $p_i$ and some polynomials $t_i$:
\begin{equation*}
    \textnormal{Ideal}(P)=\left\{\sum_{i=1}^m t_i p_i:  t_i\in\mathbb{R}[x_1,\ldots,x_n],~i=1,\ldots,m \right \}.
\end{equation*}
\end{definition}
At this point, the Krivine--Stengle Positivstellensatz can be expressed as follows:
\begin{theorem}{(Krivine--Stengle Positivstellensatz)}
Let $f_i(x)$, $i\in\mathcal{I}$, $g_j(x),\ j\in\mathcal{J}$, $h_k(x),\ k\in\mathcal{K}$ be finite sets of polynomials in $\mathbb{R}[x]$, $C=\cone(f_i)$, $M=\monoid(g_j)$, and $I=\ideal(h_k)$, then the set
\begin{equation*}
    \{x : f_i(x)\geq 0,i \in \mathcal{I},g_j(x)\neq\ 0,j \in \mathcal{J}, h_k(x)=0, k \in \mathcal{K}\}
\end{equation*}
is empty if and only if
\begin{equation*}
    \exists f\in C,\,g\in M,\,h\in I\ :\ f+g^2+h=0.
\end{equation*}
\end{theorem}

\end{document}